\documentstyle[12pt]{article}
\begin{document}
\title{Operator of Time and Generalized Schr\"odinger Equation}
\author{Slobodan Prvanovi\'c \\
{\it Scientific Computing Laboratory,}\\
{\it Center for the Study of Complex Systems,}\\
{\it Institute of Physics Belgrade, } \\
{\it University of Belgrade, Pregrevica 118, 11080 Belgrade,}\\
{\it Serbia}}
\date{}
\maketitle

\begin{abstract}

Within the framework of self-adjoint operator of time in non-relativistic quantum mechanics the equation describing change of the state of quantum 
system with respect to energy is introduced. The operator of time appears to be the generator of the change of energy while the operator of energy, 
that is conjugate to the operator of time, generates time evolution in this proposal. Two examples, one with discrete time and the other with 
continuous one, are given and the generalization of Schr\"odinger equation is proposed. 
\end{abstract}

The most important equation of quantum mechanics is, of course, the Schr\"odinger equation [1]. It is seen as equation that, for particular choice of 
Hamiltonian, determines how state of quantum mechanical system changes with respect to time. Usually, the Hamiltonian is a function of coordinate and 
momentum, $H(\hat q, \hat p )$, but there are situations in which one is interested in time dependent Hamiltonians, as well [2-8]. In these cases, one can 
find how the energy of system changes in time, which is the consequence of the influence of environment on a system under consideration. However, the 
Schr\"odinger equation is not appropriate for finding an answer to question how the state of quantum system changes with respect to the change of 
energy, acquired or lost by the system. In order to tackle this problem, one needs equation in which, so to say, one differentiates with respect to the energy, 
not with respect to the time as in the Schr\"odinger equation. This is what we are going to discuss in the present article and this will be done by using 
the formalism of the operator of time. 

There is a whole variety of topics and approaches related to the operator of time, {\it e.g.}, [9-11] and references therein, but let us shortly review our 
approach that we have proposed in [12-14].

In order to fulfill demand coming from general relativity, that space and time should be treated on equal footing, just like for every spatial degree of 
freedom a separate Hilbert space is introduced, one should introduce Hilbert space where operator of time $\hat t$ acts. More concretely, for the 
case of one degree of freedom, beside $\hat q$ and conjugate momentum $\hat p$, acting non-trivially in ${\cal H}_q$, there should be ${\cal H}_t$ where 
$\hat t$, together with $\hat s$ that is conjugate to $\hat t$, act non-trivially. So, in ${\cal H}_q \otimes {\cal H}_t$ for the self-adjoint operators 
$\hat q \otimes \hat I$, $\hat p \otimes \hat I$, $\hat I \otimes \hat t$ and $\hat I \otimes \hat s$ the following commutation relations hold:
$$
{1\over i\hbar} [\hat q \otimes \hat I, \hat p \otimes \hat I ] = \hat I \otimes \hat I ,
$$
$$
{1\over i\hbar} [\hat I \otimes \hat t , \hat I \otimes \hat s ] = - \hat I \otimes \hat I .
$$
The other commutators vanish. The operator of time $\hat t$ has continuous spectrum $\{ -\infty , +\infty \}$, just like the operators of coordinate 
and momentum $\hat q$ and $\hat p$. So is the case for the operator $\hat s$, that is conjugate to time, and which is the operator of energy. After 
noticing complete similarity among coordinate and momentum on one side and time and energy on the other side, one can introduce eigenvectors of $\hat t$: 
$$
\hat t \vert t \rangle = t \vert t \rangle, \ \ \ \ for \ \ every \ \ t \in {\bf R}.
$$
In $\vert t \rangle$ representation, operator of energy becomes $i \hbar {\partial \over \partial t}$, while its eigenvectors $\vert E \rangle$ 
become $e^{{1\over i\hbar} Et}$, for every $E\in \bf R$.

After Pauli, it is well known that there is no self-adjoint operator of time that is conjugate to the Hamiltonian $H (\hat q, \hat p )$ which has 
bounded from below spectrum. Within our proposal, self-adjoint operator of time is conjugate to the operator of energy, which has unbounded spectrum. 
The Hamiltonian and the operator of energy are acting in different Hilbert spaces, but there is subspace of the total Hilbert space where:
\begin{equation}
\hat s \vert \psi \rangle = H (\hat q, \hat p ) \vert \psi \rangle  .
\end{equation}
The states that satisfy this equation are physical since these states have non-negative energy. The last equation is nothing else but the 
Schr\"odinger equation. By taking $\vert q \rangle \otimes \vert t \rangle$ representation of previous equation, one gets the familiar form 
of Schr\"odinger equation:
\begin{equation}
i \hbar {\partial \over \partial t } \psi (q,t) = \hat H \psi (q,t) ,
\end{equation}
with the shorthand notation $\hat H = \langle q \vert H (\hat q , \hat p ) \vert q' \rangle$. In other words, operator of energy has negative 
eigenvalues as well as non-negative, but the Schr\"odinger equation appears as a constraint that selects physically meaningful states, {\it i. e.}, 
states with non-negative energy, due to the non-negative spectrum of $H (\hat q , \hat p )$, are selected by the Schr\"odinger equation. 
For the time independent Hamiltonian, typical solution of the Schr\"odinger equation (2) is $\psi _E (q) e^{{1\over i \hbar} Et}$, which is 
$\vert q \rangle \otimes \vert t \rangle$ representation of $\vert \psi _E \rangle \otimes \vert E \rangle $, where $H (\hat q , \hat p ) \vert 
\psi _E \rangle = E \vert \psi _E \rangle$ and $\hat s \vert E \rangle = E \vert E \rangle$. For such states, the Heisenberg uncertainty relation 
for $\hat s$ and $\hat t$ obviously holds. The energy eigenvectors $\vert E \rangle$ have the same formal characteristics as, say, the momentum 
eigenvectors, {\it i.e.}, they are normalized to $\delta (0)$ and, for different values of energy, they are mutually orthogonal. 

Therefore, the Schr\"odinger equation demands equal action of the operator of energy $\hat s$ and Hamiltonian $H(\hat q , \hat p)$ on states of quantum 
system and, on the other hand, due to the fact that the operator of energy in time representation is $i\hbar {\partial \over \partial t}$, it is said  
for the Schr\"odinger equation that it describes how the state of a quantum system changes with time. In the Schr\"odinger equation the Hamiltonian 
appears as dynamical counterpart of energy. The Hamiltonian is often seen as the one that determines time evolution of states of quantum system since 
its eigevalues appear in phase factor $\langle t \vert  E \rangle = e^{{1\over i \hbar} Et}$. 

In analogy with this, one can introduce:
\begin{equation}
\hat t \vert \psi \rangle = G (\hat q, \hat p ) \vert \psi \rangle  .
\end{equation}
By taking $\vert p \rangle \otimes \vert E \rangle$ representation of previous equation, one gets:
\begin{equation}
- i \hbar {\partial \over \partial E } \psi (p,E) = \hat G \psi (p,E), 
\end{equation}
With the shorthand notation $\hat G = \langle p \vert G (\hat q , \hat p ) \vert p' \rangle$. (Instead of $\vert p \rangle$ representation one can, 
of course, use $\vert q \rangle$ representation. The momentum representation is taken just because energy and momentum form quadri vector. The similar 
remark applies for the above $\vert q \rangle \otimes \vert t \rangle$ representation.) 

In (3) one demands that the operator of time and its dynamical counterpart $G (\hat q, \hat p )$ should equally act on states of quantum mechanical 
system. As the original Schr\"odinger equation, this equation represents constraint, as well. The original Schr\"odinger equation can be seen as the 
energy constraint, while this one is time constraint that also selects subspace in ${\cal H}_q \otimes {\cal H}_t$. After representing (3) in 
$\vert p \rangle \otimes \vert E \rangle$ basis, in analogy with the interpretation of original Schr\"odinger equation, it could be said that this equation 
describes how the state of quantum mechanical system changes with respect to the change of energy. Just like the time is treated as evolution parameter 
for (2), the energy in (4) can be seen as independent variable. 
 
The typical solution of, let us call it, the second Schr\"odinger equation (4) is $\psi _ t (p) e^{{-1\over i \hbar} Et}$. It is $\vert p \rangle \otimes 
\vert E \rangle$ representation of $\vert \psi _t \rangle \otimes \vert t \rangle $, where $G (\hat q , \hat p ) \vert \psi _t \rangle = t \vert \psi _t 
\rangle$ and $\hat t \vert t \rangle = t \vert t \rangle$. For such states, the Heisenberg uncertainty relation for $\hat s$ and $\hat t$ obviously holds. 

Let us stress that the generators of the Lie algebra (beside coordinate and momentum) are $\hat s$ and $\hat t$, and not $H(\hat q , \hat p)$ and 
$G(\hat q , \hat p )$. For the former it holds $ {1\over i\hbar} [\hat t , \hat s ] = - \hat I $, while for the latter ones, since they are just 
dynamical counterparts of the energy and time, {\it a priori} there is no reason to demand non-commutativity, {\it i. e}, their commutator depends 
on particular choice of these functions of coordinate and momentum. 

As for different systems are appropriate different Hamiltonians, for different situations one should use appropriate $G (\hat q, \hat p )$, As an 
illustration, let us briefly discuss two examples of $G(\hat q , \hat p )$. Due to the Big Bang as the beginning of time, it seems reasonable to assume 
$G(\hat q , \hat p )$ with bounded from below spectrum. Without going into debate about existence of time crystals, the first example offers a toy model of 
discrete time while for the second one the quantum system is characterized with the continuous time.

If $G(\hat q , \hat p )$ is:
$$
G(\hat q , \hat p )= {\hbar \over m^2 c^4} ({\hat p^2 \over 2m} + {1\over 2} m \omega \hat q^2),
$$
then solutions of the second Schr\"odinger equation (3) are $\vert \psi _{n} \rangle \otimes \vert t_n \rangle $, $n \in {\bf {\rm N}}$, where $\vert \psi _{n} 
\rangle $ are well known solutions of the eigenvalue problem for Hamiltonian of harmonic oscillator and:
$$
t_n = {\hbar ^2 \omega \over m^2 c^4} (n + {1\over 2}).
$$ 

If the standard ladder operators $\hat a^{\dagger}$ and $\hat a$, that act on $\vert \psi _{t_n} \rangle $, are directly multiplied by time translation operator 
$\hat a^{\dagger} \otimes e^{{1\over i \hbar} \hat s \Delta t}$ and $\hat a \otimes e^{{1\over i \hbar} \hat s \Delta t}$, where $\Delta t = {\hbar ^2 \omega 
\over m^2 c^4}$, then it holds: 
$$
\hat a^{\dagger} \otimes e^{-{1\over i \hbar} \hat s \Delta t} \vert \psi _{t_n} \rangle \otimes \vert t_n \rangle ={\sqrt {n+1}} \vert \psi _{t_{n+1}} \rangle 
\otimes \vert t_{n+1} \rangle ,
$$
$$
\hat a \otimes e^{{1\over i \hbar} \hat s \Delta t} \vert \psi _{t_n} \rangle \otimes \vert t_n \rangle ={\sqrt {n}} \vert \psi _{t_{n-1}} \rangle 
\otimes \vert t_{n-1} \rangle .
$$
So, these operators formally describe transition between states that represent nearby moments in time. Distinguished moments $t_n$ are the ones in which system 
can be located in time. 

The unitary operator attached to change of the energy is $e^{{1\over i \hbar} \hat t \Delta E}$. {\it Nota bene}, the unitary 
operator that represents time evolution is $e^{{1\over i \hbar} \hat s \Delta t}$, not the $e^{{1\over i \hbar} H(\hat q , \hat p ) \Delta t}$, because 
$\hat s$ is the generator of time translation in ${\cal H }_t$, and not $ H(\hat q , \hat p ) $ since it does not even act in ${\cal H }_t$. Of course, 
these two unitary operators are effectively the same when they act on solutions of the original Schr\"odinger equation. 

Another example is the case with continuous time, when $G(\hat q , \hat p )= {\hbar \over m^3 c^4}\hat p^2$. The solutions of the second Schr\"odinger equation 
are then $\vert p \rangle \otimes \vert t \rangle $, where $t = {\hbar \over m^3 c^4}p^2$. This $G(\hat q , \hat p )$ commutes with the Hamiltonian of a free 
particle, while the previous one obviously commutes with the Hamiltonian of harmonic oscillator. Needless to say, in general, one can combine every $H(\hat q , 
\hat p )$ with any $G(\hat q , \hat p )$. However, with particular $H(\hat q , \hat p )$ and $G(\hat q , \hat p )$ one defines some physical system for which, 
according to the standard interpretation, it is possible to describe changes in time and energy. By staying with the standard interpretation, one can say that 
if some quantum system evolves under the action of $H(\hat q , \hat p )$ until the moment, say, $t_0$, when it changes energy, then one should apply 
$\hat U (i,j) \otimes e^{{1\over i \hbar} \hat t \Delta E}$ on state $\vert E_i \rangle \otimes \vert E_i \rangle$, that is solution of the original Schr\"odinger 
equation for $E_i$, and get the other solution $\vert E_j \rangle \otimes \vert E_j \rangle$ of the same equation (here $\hat U (i,j)$ is the unitary operator 
that rotates $\vert E_i \rangle$ into $\vert E_j\rangle$ in the first Hilbert space and $\Delta E$ is the energy difference among given states). 
The moment $t_0$ at which this can happen depends on $G(\hat q , \hat p )$, {\it i. e.}, it has to belong to the spectrum of $G(\hat q , \hat p )$, just like $E_i$ 
belongs to the spectrum of $H(\hat q , \hat p )$. In other words, $H(\hat q , \hat p )$ and $G(\hat q , \hat p )$ determine what are the solutions of original 
and second Schr\"odinger equations, and everything related to this. In particular, on $H(\hat q , \hat p )$ and $G(\hat q , \hat p )$ depend in which amounts energy 
of the system can be changed and in what time intervals this can happen if there is a sequence of time evolutions followed by the changes of the systems energy. 

This leads us to the generalization of the original Schr\"odinger equation:
\begin{equation}
(c_s \hat s + c_t \hat t ) \vert \Psi \rangle = F (\hat q , \hat p , \hat s , \hat t ) \vert \Psi \rangle , \ \ \ \ \ \vert \Psi \rangle \in {\cal H}_q 
\otimes {\cal H}_t .
\end{equation}
This equation superposes both types of the systems change. Obviously, the original and second Schr\"odinger equation follow from (5) for the adequate choices 
of dimensional constants $c_s$, $c_t$ and $F (\hat q , \hat p , \hat s , \hat t )$. This equation is appropriate for investigations of situations with time 
dependant Hamiltonians, as well. However, such Hamiltonians shall be considered elsewhere. 

Finally, let us briefly comment measurement outcomes of energy and time. Within the present formalism, both $\hat s$ and $\hat t$ have the whole ${\bf {\rm R}}$ as 
spectrum, but only the values determined by $H(\hat q , \hat p )$ and $G(\hat q , \hat p )$ through Schr\"odinger equations can be attributed to the system under 
consideration. In other words, the physical system is defined by $H(\hat q , \hat p )$ and $G(\hat q , \hat p )$ and Schr\"odinger equations select values of 
energy and time that are characteristic to that system. While other values are formally possible, they are unrelated to the system. Therefore, description of 
measurement process should use orthogonal resolution of identity operator in a subspace determined by (1) or (3), not the identity operator in whole 
${\cal H}_q \otimes {\cal H}_t$.

\section{Acknowledgement}
We acknowledge support of the the Serbian Ministry of education, science and technological development, contract ON171017.

\end{document}